\documentclass[twocolumn,floatfix,showpacs]{revtex4}

\usepackage{graphicx}
\usepackage{amsmath}
\usepackage{amssymb}

\usepackage{color}

\begin{document}

\title{Top-transmon: hybrid superconducting qubit for parity-protected quantum computation}

\author{F. Hassler}
\affiliation{Instituut-Lorentz, Universiteit Leiden, P.O. Box 9506, 2300 RA Leiden, The Netherlands}
\author{A. R. Akhmerov}
\affiliation{Instituut-Lorentz, Universiteit Leiden, P.O. Box 9506, 2300 RA Leiden, The Netherlands}
\author{C. W. J. Beenakker}
\affiliation{Instituut-Lorentz, Universiteit Leiden, P.O. Box 9506, 2300 RA Leiden, The Netherlands}

\date{May 2011}

\begin{abstract}
Qubits constructed from uncoupled Majorana fermions are protected from
decoherence, but to perform a quantum computation this topological
protection needs to be broken. Parity-protected quantum computation
breaks the protection in a minimally invasive way, by coupling directly
to the fermion parity of the system --- irrespective of any quasiparticle
excitations. Here we propose to use a superconducting charge qubit in a transmission line resonator 
(a socalled \textit{transmon})
to perform parity-protected rotations and read-out of a topological (\textit{top}) qubit. The
advantage over an earlier proposal using a flux qubit is that the coupling
can be switched on and off with exponential accuracy, promising a reduced
sensitivity to charge noise.
\end{abstract}


\maketitle

\section{Introduction}
\label{intro}

Condensed matter systems with quasiparticles that are Majorana fermions
(equal to their own antiparticle) offer a promising platform for topological
quantum computation \cite{Kit03,nayak:08}. Majorana fermions are non-Abelian
anyons of the Ising type \cite{read:00,ivanov:01}, for which topologically
protected operations (braiding) are insufficient to implement a universal
quantum computation. Bravyi and Kitaev \cite{Bra05} showed that two types
of phase coherent operations without topological protection are needed in
addition to braiding: single-qubit rotations and joint read out of up to two qubits.

There is a great variety of proposals how to implement
the unprotected operations without losing phase coherence
\cite{Fu08,Akh09,Has10,Sau10a,Fu10,Cla10,flensberg,Jia11,Gro10}. What
most of these proposals have in common, is that they are sensitive both to
decoherence of quasiparticle excitations and to charge noise. The latter
is required to jointly read out two qubits \cite{note1},
but the former can be avoided. Parity-protected quantum computation (PPQC)
\cite{Has10} relies on the coherent manipulation of the charge degree of
freedom of small superconducting islands, without requiring quasiparticle
coherence. Insensitivity to quasiparticle decoherence allows to perform
quantum computations even when Majorana fermions coexist with thermally
excited states \cite{Akh10}, which typically have a very small excitation
gap \cite{Car64, Sel11} (although there exist ways to increase that gap
\cite{Sau10b,Mao10,Pot11}).

The specific proposal for PPQC introduced in Ref.~\cite{Has10} is to use the
Aharonov-Casher effect \cite{stern:08,Gro11} to couple Majorana fermions in a nanowire 
\cite{lutchyn:10,oreg:10} to a flux qubit~\cite{Moo99}. The flux qubit
is a nontopological superconducting qubit, which can reliably be rotated
and read out by microwaves. The Aharonov-Casher effect couples to the charge
modulo $2e$ of the nanowire. This coupling is insensitive
to sub-gap excitations or Cooper pair tunneling events through Josephson
junctions, which do not change the fermion parity of the nanowire. The remaining
sensitivity to charge noise is minimized by decoupling the flux qubit from
the topological qubit when the operation is completed.

In this paper we consider an alternative way to perform parity-protected
operations on a topological qubit, with an expected reduced sensitivity
to charge noise. Instead of a flux qubit, we propose to use the socalled
transmon qubit, which is a superconducting charge qubit with large ratio of
Josephson energy over charging energy \cite{koch:07,Hou09,dicarlo:09}. 
(The transmon is placed in a transmission line resonator for read out, hence the name.)
The transmon
and flux qubit both couple to the fermion parity of the topological qubit,
but while the coupling strength of the flux qubit can only be varied as
a power law in the magnetic field, this variation is exponential in the
field strength for the transmon qubit \cite{schreier:08}.

In the next section we introduce this hybrid topological-transmon
qubit (abbreviated to {\em top-transmon}) by discussing the two building
blocks separately. We show how the coupling can be switched on and off
exponentially by tuning the magnetic field. Then, we outline how the coupling
can be used to rotate the topological qubit, as well as to jointly
read out sets of topological qubits. Together with braiding, these
are the operations required for a universal quantum computer \cite{Bra05}.

We also discuss how the top-transmon permits quantum
state transfer between topological and non-topological qubits. This is
an alternative to earlier proposals using a superconducting flux qubit
\cite{Jia11} or normal-state charge qubit \cite{bonderson:10} as the
non-topological qubit (which lacked parity protection or the possibility
to switch the coupling off exponentially). At the end of the paper we give an estimate
of the relevant time scales for these parity-protected operations.

\section{Top-transmon}

\begin{figure}[tb]
  \centering
  \includegraphics[width=0.8\linewidth]{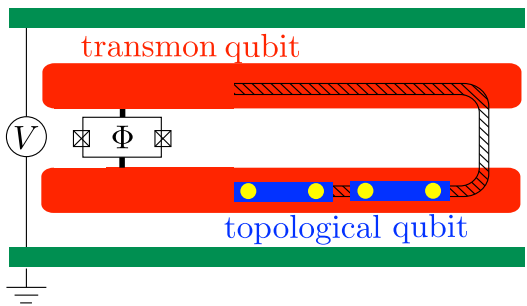}
  \caption{%
  Schematic design of the top-transmon. Two superconducting islands (red), connected by a split Josephson junction (crosses), and placed in a superconducting transmission line resonator (green) form the transmon qubit. A voltage $V$ controls the induced charge on the islands and the magnetic fiux $\Phi$ enclosed by the Josephson junction controls the  charge sensitivity of the transmon. The topological qubit is formed by two pairs of Majorana fermions (yellow dots), at the end points of two undepleted segments (blue) of a semiconductor nanowire (shaded ribbon indicates the depleted region). To read out and rotate the topological qubit, one pair of Majorana fermions is moved onto the other island.
  }\label{fig:circuit}
\end{figure}

We consider a pair of superconducting islands coupled through a Josephson junction. A semiconductor nanowire with strong spin-orbit coupling (typically InAs) is placed on the islands. The superconducting proximity effect in a parallel magnetic field can produce midgap states at the end points of undepleted sections of the nanowire  \cite{lutchyn:10,oreg:10}. Each midgap state binds a Majorana fermion. The transmon qubit is formed by the superconducting islands, while the topological qubit is formed by sets of four Majorana bound states. A schematic diagram of this hybrid ``top-transmon'' qubit is shown in Fig.\ \ref{fig:circuit}. We introduce the two building blocks in separate subsections.

\subsection{Transmon qubit}

The superconducting islands have Hamiltonian $H= H_C + H_J$,
containing the capacitive energy due to a charge difference
of the islands and the Josephson energy due to the tunneling
of Cooper pairs between the islands \cite{makhlin:01,Dev04}.

The capacitive energy (for capacitance $C$) is given by
\begin{equation}
H_C = \frac{(2e\delta N - q_\text{ind})^2}{2 C}.\label{eq:hamc}
\end{equation}
The difference $\delta N=(N_{1}-N_{2})/2$ of the number of Cooper pairs $N_{i}$ on each island changes by unity at each pair tunnel event. The induced (or offset) charge $q_{\rm ind}$ remains constant during pair tunnel events, but can be varied externally by a gate voltage $V$ (coupled to the islands via a capacitance $C_{g}$). 

Unpaired electrons ($n_{i}$ on each island) also contribute to $q_{\rm ind}$,
\begin{equation}
q_{\rm ind}=C_{g}V-(\tfrac{1}{2}-f_{C})en_{1}+(\tfrac{1}{2}+f_{C})en_{2}+2ef_{C}(N_{1}+N_{2}).\label{qinddef0}
\end{equation}
The coefficient $f_{C}=0$ if the capacitance matrix has $1\leftrightarrow 2$ exchange symmetry, but is nonzero for an asymmetric structure. (See App.\ \ref{paritymeter} for a calculation.) Regardless of the value of $f_{C}$, the induced charge changes by $\pm e$ if an unpaired electron is transported from one island to the other.

The Josephson energy is
\begin{equation}\label{eq:hamj}
  H_J = - E_J \cos \varphi,
\end{equation}
with $\varphi$ the (gauge-invariant) phase difference across the Josephson junction. We need a tunable Josephson energy $E_{J}$, which can be achieved
by replacing the single Josephson junction by a pair of identical Josephson junctions in parallel \cite{makhlin:01,Dev04}. 
The Josephson energy then depends on the 
magnetic flux $\Phi$ enclosed by the two Josephson junctions (each with coupling energy $E_{0}$),
\begin{equation}
E_J(\Phi) = 2 E_{0} \cos(\pi \Phi/\Phi_0).
\end{equation}
The Josephson coupling is maximal when $\Phi$ is an integer multiple of
the superconducting flux quantum $\Phi_0 = h/2e$.

In a quantum mechanical description the number $\delta N$ and phase $\varphi$ 
are conjugate operators, hence their commutator
$[\delta N,\phi]=-i$ and in the phase basis $\delta N=-i\partial/\partial\varphi$.
In terms of the ladder operators $n^{\pm}=e^{\pm i\varphi}$ the Josephson
energy takes the form of a tunneling Hamiltonian,
\begin{equation}\label{eq:tun}
  H_J = -\tfrac{1}{2}E_J (n^+ + n^-),
\end{equation}
where $n^+$ tunnels a Cooper pair from island 1 to island 2 and $n^- = (n^+)^\dag$ 
describes the reverse process. These processes govern the (differential) Cooper pair box \cite{Sch07}.

Typically \cite{Dev04}, in a Cooper pair box the charging energy $E_C =
e^2/2C$ is much larger than the Josephson energy $E_J$.
The spectrum is then given approximately by the eigenvalues of $H_{C}$,
\begin{equation}\label{eq:spectrum}
  E_n =4 E_C \Bigl(n - \frac{q_\text{ind}}{2e} \Bigr)^2,\;\;{\rm if}\;\;E_{C}\gg E_{J}.
\end{equation}
The role of the Josephson energy is to remove degeneracies,
for example, the crossing of
$E_1$ and $E_0$ at $q_\text{ind} = e$ becomes an anticrossing with a gap $E_{J}$.
The spectrum as a function of $q_\text{ind}$ thus consists of bands with periodicity $2e$.
Close to the anticrossing of $E_{1}$ and $E_{0}$ the Hamiltonian can be approximated by a two-level system,
with Hamiltonian
\begin{equation}\label{eq:qubit}
  H_\text{qubit} = 2 E_C (q_\text{ind}/e - 1) \sigma_z 
  - \tfrac{1}{2}E_J \sigma_x.
\end{equation}
This superconducting charge qubit \cite{Nak99,vion:02}
is deficient because of its sensitivity to fluctuations in 
$q_{\rm ind}$ (charge noise). 

\begin{figure}[tb]
  \centering
  \includegraphics[width=\linewidth]{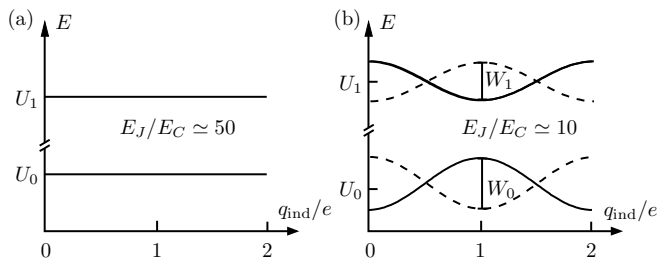}
  \caption{%
  Energy spectrum of the transmon qubit as a function of the induced charge
  for the two lowest energy states, at two values of the ratio $E_{J}/E_{C}$ of
  Josephson energy and charging energy. (Solid and dashed curves differ by an additional unpaired electron.)
  The sensitivity to charge noise
  can made exponentially small by increasing $E_{J}/E_{C}$ [going from (b) to (a)].
  }\label{fig:spect}
\end{figure}

The transmon \cite{koch:07,Hou09,dicarlo:09} removes the deficiency
by operating in the regime $E_J \gg E_C$. The band index $n$ then no longer specifies the charge on the islands, 
since Cooper pairs can tunnel freely through the Josephson junction.
The dependence of $E_n$ on $q_{\rm ind}$ now has approximately a cosine form (see Fig.~\ref{fig:spect}),
\begin{equation}\label{eq:band}
  E_n = U_n - (-1)^{n}\frac{W_n}{2} \cos(\pi q_\text{ind}/e),\;\;{\rm if}\;\;E_{J}\gg E_{C},
\end{equation}
with band width $W_n \propto \exp( - \sqrt{8 E_J/E_C})$ and band
spacing $U_{n+1} - U_n \simeq \sqrt{E_C E_J}$. The sensitivity to charge noise can be made exponentially 
small by increasing $E_{J}$ (through a variation of the flux $\Phi$),
so that the bands become flat and fluctuations in $q_{\rm ind}$ do not lead to an uncertainty in energy (which is the origin of dephasing).

In a typical device \cite{Hou09}, the band spacing $(E_{1}-E_{0})/h$ varies in the range $3-5\,{\rm GHz}$, for ratios $E_{J}/E_{C}$ increasing from $10$ to $30$. The band width, in contrast, drops by two orders of magnitude from $W/h\simeq 100\,{\rm MHz}$ to $1\,{\rm MHz}$.

The state of the transmon qubit can be read out by sending a microwave probe beam through the transmission line resonator \cite{wallraff:04,schuster:05,schreier:08}. The resonance frequency $\omega_{\rm res}$ depends on whether the transmon is in the ground state ($|E_{0}\rangle$, $s_{z}=-1$) or in the first excited state ($|E_{1}\rangle$, $s_{z}=+1$), according to
\begin{equation}
\omega_{\rm res}=\omega_{0}-\frac{s_{z} g^{2}}{\omega_{0}-(E_{1}-E_{0})/\hbar}.
\end{equation}
Here $\omega_{0}/2\pi\simeq 10\,{\rm GHz}$ is the bare resonance frequency and $g/2\pi\simeq 100\,{\rm MHz}$ the qubit-resonator coupling strength. The detuning $\omega_{0}-(E_{1}-E_{0})/\hbar$ is large compared to $g$, so that the resonance is only slightly shifted. Such small shifts can be measured sensitively as a phase shift of the transmitted microwaves.

\subsection{Topological qubit}

A pair of Majorana bound states forms a two-level system and hence a qubit. The fermion parity is even in the lower state $|0\rangle$ and odd in the upper state $|1\rangle$. A logical qubit is constructed from two of these fundamental qubits \cite{nayak:08}, hence from four Majorana bound states. Without loss of generality we may assume that the combined fermion parity is even. The state of the logical qubit is then given by
\begin{equation}
|\psi\rangle=\alpha|00\rangle+\beta|11\rangle,\;\;|\alpha|^{2}+|\beta|^{2}=1.\label{psidef}
\end{equation}

The read-out operation projects $|\psi\rangle$ on the state $|00\rangle$ or $|11\rangle$. This is a fermion parity measurement of one of the two fundamental qubits that encode the logical qubit  \cite{Fu10}. For a universal quantum computation it is also needed to perform a joint parity measurement on one fundamental qubit from the logical qubit plus one additional ancilla qubit \cite{Bra05}. Hence the required parity measurements will involve either two or four Majorana bound states. We denote this fermion parity by $n_{M}\in\{0,1\}$.

The Majorana bound states are located at the end points of undepleted segments of the nanowire, initially all on one of the two superconducting islands. The voltage $V$ is adjusted so that initially $q_{\rm ind}=0$ (modulo $2e$). Gate electrodes (not shown in Fig.\ \ref{fig:circuit}) transport onto the other island the Majorana bound states that are to be measured. The induced charge changes as a result of this operation, $q_{\rm ind}\mapsto en_{M}$ (modulo $2e$), and so directly couples to the required fermion parity.

\section{Parity-protected operations}

In this section we show how the transmon qubit can be used to perform parity-protected operations on the topological qubit. Topologically protected braiding operations can be performed by means of T-junctions of nanowires \cite{alicea:10,Cla10b}, and will not be considered here.

Common to all parity-protected operations is that the flux $\Phi$ through the Josephson junction is kept close to zero (modulo $\Phi_{0}$) both before and after the operation. The ratio $E_{J}(\Phi)/E_{C}$ is then much larger than unity, hence the coupling between transmon and topological qubit is exponentially small. During the operation $\Phi$ is adjusted to a value close to $\Phi_{0}/2$ (modulo $\Phi_{0}$), so that the transmon becomes sensitive to the fermion parity of the topological qubit. 

\subsection{Read out and phase gate}

The operations of read out and phase gate (= single-qubit rotation) proceed in the same way as in Ref.\ \cite{Has10}, where the topological qubit was coupled to a flux qubit rather than to a transmon (with the key difference that there the coupling could not be switched off exponentially). We summarize the procedure.

To read out the topological qubit, two of the four Majorana fermions that encode the logical qubit are moved from one island to the other. (A joint parity measurement on two topological qubits can likewise be performed by moving four Majorana fermions to the other island.) Depending on the fermion parity $n_{M}$, the level spacing $\Delta E=E_{1}-E_{0}$ of the transmon qubit is given by
\begin{equation}
\Delta E=\left\{\begin{array}{ll}
U_{1}-U_{0}+(W_{0}+W_{1})/2,&{\rm if}\;\;n_{M}=0,\\
U_{1}-U_{0}-(W_{0}+W_{1})/2,&{\rm if}\;\;n_{M}=1,
\end{array}\right.
\label{DeltaEdef}
\end{equation}
in view of Eq.\ \eqref{eq:band} with $q_{\rm ind}=en_{M}$ (modulo $2e$). A measurement of $\Delta E$ by microwave spectroscopy thus determines $n_{M}$.

The read-out operation projects the state $|\psi\rangle$ in Eq.\ \eqref{psidef} onto either $|00\rangle$ or $|11\rangle$. A single-qubit rotation is a unitary operation, rather than a projective measurement. For that purpose one would couple the transmon to the topological qubit without microwave irradiation. The transmon is initialized in the ground state $|E_{0}\rangle$. In a time $\tau$ the coupled topological qubit evolves as
\begin{align}
\alpha |00\rangle+\beta|11\rangle\mapsto{}& \alpha e^{i(U_{0}-W_{0}/2)\tau/\hbar}|00\rangle\nonumber\\
&+\beta e^{i(U_{0}+W_{0}/2)\tau/\hbar}|11\rangle.\label{phasegate}
\end{align}
(The transmon stays in the ground state during the operation, so it factors out of the wave function.) If the coupling time is chosen such that $\tau W_{0}/2\hbar = \theta$ one performs a $\theta$-phase gate operation.

Eq.\ \eqref{phasegate} amounts to a rotation of the logical qubit by an angle $2\theta$. While $\pi/2$ rotations can be performed by braiding, other rotation angles require breaking of the topological protection. For a universal quantum computation a $\pi/4$ rotation is sufficient \cite{Bra05}, obtained from Eq.\ \eqref{phasegate} by choosing $\theta=\pi/8$.

\subsection{Quantum state transfer}

The operation of quantum state transfer starts from the topological qubit in the state $|00\rangle$ and the transmon in an arbitrary \textit{unknown} superposition $\alpha|E_{0}\rangle+\beta|E_{1}\rangle$ of ground state and first excited state. At the end of the operation the topological qubit is in the state $\alpha|00\rangle+\beta|11\rangle$, while the transmon is no longer in a superposition state. The reverse operation is also possible (transfer of an unknown state from the topological qubit to the transmon).

As we will show in this subsection, quantum state transfer can be performed by a combination of topologically protected braiding operations, plus a parity-protected {\sc cnot} operation on the top-transmon. The procedure is an alternative to the quantum state transfer proposals of Refs.\ \cite{Jia11,bonderson:10}, which coupled the topological qubit to a flux qubit. By using a transmon rather than a flux qubit, we can offer both parity protection (no sensitivity to quasiparticle excitations) and the ability to switch the sensitivity to charge noise off in an exponential manner.

The {\sc cnot} operation that we need is a conditional $\sigma_{x}$ operation on the transmon, where the condition for the $\sigma_{x}$ operation (so a switch $|E_{0}\rangle\leftrightarrow|E_{1}\rangle$) is that the topological qubit is in the state $|11\rangle$. This is a {\sc cnot} gate for the top-transmon, with the topological qubit as the control and the transmon as the target.

The $\sigma_{x}$ operation on the transmon is a $\pi$-pulse of microwave radiation, at the resonant frequency $\Delta E/\hbar$. As before, two of the four Majorana fermions that encode the topological qubit are moved from one island to the other. The resonant frequency then depends on their fermion parity $n_{M}$ through Eq.\ \eqref{DeltaEdef}. If the microwave radiation is resonant for $n_{M}=1$, then for $n_{M}=0$ it is detuned by an amount $W_{0}+W_{1}$. The $\pi$-pulse therefore performs the required {\sc cnot} operation if the detuning is larger than the resonance width, which is the Rabi frequency $\Omega_{\rm Rabi}$.

\begin{figure}[!t]
  \centering
  \includegraphics{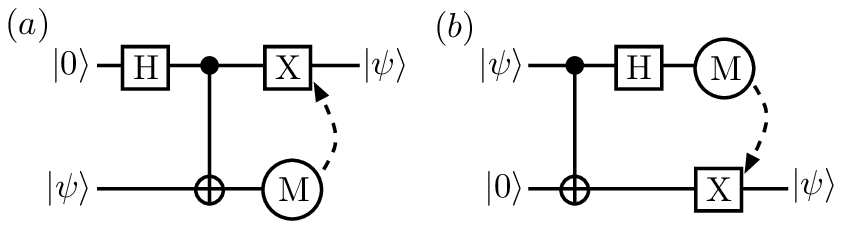}
  \caption{%
  Diagram for quantum state transfer in the top-transmon, starting from an unknown state $|\psi\rangle$ in the transmon qubit (a) or in the topological qubit (b). Adapted from Ref.\ \cite{nielsen} and explained further in the text.
  }\label{fig:swap}
\end{figure}

The two diagrams in Fig.\ \ref{fig:swap} show, following Ref.\ \cite{nielsen}, how quantum state transfer (in both directions) is achieved with the help of the {\sc cnot} gate. Two more operations are needed, a Hadamard gate $H=(\sigma_{x}+\sigma_{z})/\sqrt{2}$ on the topological qubit and a {\sc not} gate $X=\sigma_{x}$ on one of the qubits --- conditionally on the outcome of the measurement (M) of the other qubit. (Both Hadamard and {\sc not} gates can be performed on the topological qubit by braiding.)

We explain diagram (a): The topological qubit (upper line) starts out in the state $|00\rangle$ and is transformed by the Hadamard gate into the superposition state $(|00\rangle+|11\rangle)/\sqrt{2}$. The transmon qubit (lower line) starts out in the (unknown) state $|\psi\rangle=\alpha|E_{0}\rangle+\beta|E_{1}\rangle$.
The subsequent \textsc{cnot}-gate entangles the
two qubits, producing the state
\begin{align}
|\Psi\rangle ={}&
(|00 \rangle|\psi\rangle + | 11 \rangle \sigma_{x}|\psi\rangle )/\sqrt{2}\nonumber\\
={}&(\alpha|00\rangle+\beta|11\rangle)|E_{0}\rangle/\sqrt{2}\nonumber\\
&+(\beta|00\rangle+\alpha|11\rangle)|E_{1}\rangle/\sqrt{2}.
\end{align}
We now measure the transmon by probing the microwave resonator. If the transmon is in the ground state $|E_{0}\rangle$, no further operation is required. If it is in the excited state $|E_{1}\rangle$, then a final {\sc not} operation on the topological qubit completes the quantum state transfer.

\subsection{Time scales for parity protection}

The operation time of a $\pi/8$ phase gate is $\tau=\pi\hbar/4W_{0}$, which for $W_{0}/h\simeq 100\,{\rm MHz}$ corresponds to $\tau\simeq 1.3\,{\rm ns}$. For quantum state transfer the {\sc cnot} gate needs a $\pi$-pulse of duration $\pi/\Omega_{\rm Rabi}$. The Rabi frequency should be small compared to $2W_0/h \simeq 200\,{\rm MHz}$. Choosing
$\Omega_{\rm Rabi}/2\pi = 10\,{\rm MHz}$, the operation time of the \textsc{cnot} gate is $50\,{\rm  ns}$. These operation times are much shorter than the coherence time of the transmon qubit, which is in the $\mu{\rm s}$ range \cite{Hou09}.

The fundamental limitation to parity protection is the incoherent tunneling of unpaired electrons between the superconducting islands, a process called ``quasiparticle poisoning''. Since these tunnel events change the fermion parity, they break the parity protection of the operations. One would expect the density of unpaired quasiparticles to vanish exponentially as the temperature drops below the critical temperature of the superconductor, but experimentally a saturation is observed \cite{Mar09}. Still, the characteristic time scale for quasiparticle number fluctuations, which sets the upper limit for parity protection, becomes sufficiently large: Ref.\ \cite{Vis11} finds 2~ms in Al below 160~mK.

\section{Conclusion}

In conclusion, we have shown how a superconducting charge qubit can be used to read out and rotate a topological qubit. These are the two operations that cannot be performed by braiding of Majorana fermions, but which are needed for a universal quantum computer \cite{Bra05}. Our proposal is an alternative to the read-out and rotation by means of a superconducting flux qubit \cite{Has10}. 

In both designs, the superconducting qubit functions as a fermion parity meter. The flux qubit measures the fermion parity through the Aharonov-Casher effect, while the charge qubit relies on the $2e$ periodicity of the superconducting ground state energy. Both parity meters are insensitive to subgap excitations (parity protection). 

The advantage we see for a charge qubit over a flux qubit is that the coupling to the topological qubit can be made exponentially small, by increasing the ratio of Josephson energy $E_{J}$ over charging energy $E_{C}$. A superconducting charge qubit with adjustable $E_{J}/E_{C}$ (a socalled transmon qubit \cite{koch:07}) functions as a parity meter with an exponential on-off switch. 

The hybrid design of a coupled transmon and topological qubit (top-transmon) retains the full topological protection with exponential accuracy in the off-state ($E_{J}/E_{C}\gg 1$). In the on-state ($E_{J}/E_{C}\gtrsim 1$) the qubit is sensitive to charge noise, but still protected from noise that preserves the fermion parity.

\begin{figure}[tb]
  \centering
\includegraphics[width=1\linewidth]{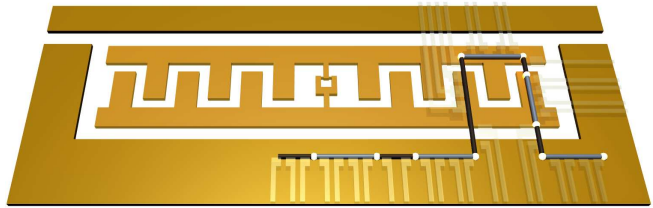}
  \caption{%
Illustration of a top-transmon. Superconductors are shown in gold, depleted segments of a nanowire are black, undepleted segments are blue, and the Majorana bound states are indicated by white dots. The thin strips on top of the superconductors are gate electrodes, used to move the Majorana's along the nanowire. The drawing shows a topological qubit consisting of four Majorana's, two on each of the superconducting islands. By adjusting the flux through the split Josephson junction connecting the islands, the transmon is coupled to the fermion parity of the topological qubit, allowing for parity-protected quantum computation. 
  }\label{fig_3D}
\end{figure}

The experimental realization of a top-transmon is a major challenge, involving a variety of design decisions that go beyond this proposal.
In Fig.\ \ref{fig_3D} we give an impression of one design, to inspire further progress in this direction.

\acknowledgments

We have benefited from discussions with L. DiCarlo and V. Manucharyan. This
research was supported by the Dutch Science Foundation NWO/FOM and by an
ERC Advanced Investigator Grant.

\appendix

\section{Differential Cooper pair box as fermion parity meter}
\label{paritymeter}

\begin{figure}[tb]
  \centering
  \includegraphics[width=0.8\linewidth]{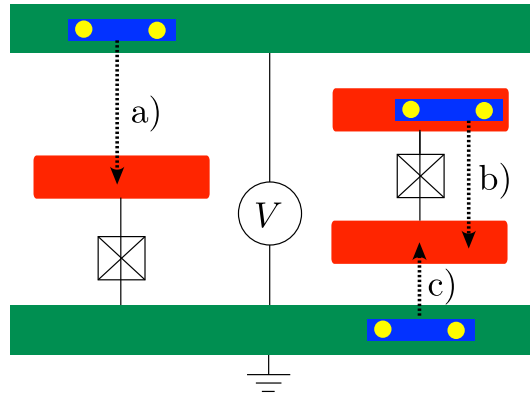}
  \caption{%
Left panel: Cooper pair box formed by a single superconducting island (red), connected to a grounded superconducting plate (green) via a Josephson junction (cross). Right panel: Differential Cooper pair box, formed by two Josephson coupled islands disconnected from ground. (For both devices, a second superconducting plate at a voltage $V$ provides control over the induced charge.) The dotted arrows indicate three pathways by which a topological qubit (yellow dots) can be brought onto a superconducting island. Pathways a) and b) allow for a fermion parity measurement, while pathway c) does not. 
  }\label{fig:CPB}
\end{figure}

The parity-protected read-out and rotation of a topological qubit requires the measurement of the charge modulo $2e$ of a nanowire containing a pair of Majorana bound states at the end points. Most directly, this measurement of the fermion parity $n_{M}\in\{0,1\}$ can be performed by bringing the nanowire in electrical contact with a Cooper pair box: a superconducting island coupled to a grounded superconductor via a Josephson junction. (See Fig.\ \ref{fig:CPB}, left panel.)

The Cooper pair box works as a parity meter because its energy spectrum $E_{n}(q_{\rm ind})$ is mapped onto $E_{n}(q_{\rm ind}+en_{M})$ when the Majorana fermions are brought on the island. If the charge $q_{\rm ind}$ induced by the gate is calibrated to zero before the operation, the spectrum $E_{n}(en_{M})$ directly determines the fermion parity.

With a single island, it is difficult to reach a large capacitance without effectively grounding the island. For that reason, the transmon uses \textit{two} superconducting islands, connected to each other via a Josephson junction and disconnected from ground \cite{koch:07}. (See Fig.\ \ref{fig:CPB}, right panel.) The spectrum of this differential Cooper pair box is not $2e$-periodic in the total charge $Q_{\rm tot}=Q_{1}+Q_{2}$ on both islands, so it cannot measure the fermion parity of external charges. If Majorana fermions are brought onto one of the islands from the outside, this operation changes $Q_{\rm tot}$ in an unknown way --- preventing the determination of the fermion parity.

To see how the $2e$-periodicity is lost, we calculate the electrostatic energy of the two islands,
\begin{equation}
H_{C}=\frac{1}{2}\sum_{i,j=1}^{2}(Q_{i}-q_{i})(Q_{j}-q_{j})C^{\rm inv}_{ij},\label{ECij}
\end{equation}
determined by the charges $Q_{i}$ on the islands, the charges $q_{i}=C_{g,i}V$ induced by the gates, and the elements $C^{\rm inv}_{ij}$ of the inverse of a (symmetric) capacitance matrix $C$. Island $i$ contains $N_{i}$ Cooper pairs plus $n_{i}$ unpaired electrons, so $Q_{i}=2eN_{i}+en_{i}$. 

We assume that $N_{\rm tot}=N_{1}+N_{2}$ is even, so that $\delta N=(N_{1}-N_{2})/2\in\mathbb{Z}$. (This is without loss of generality, since if $N_{\rm tot}$ is odd we can make it even without changing the energy by redefining $N_{1}\mapsto N_{1}-1$, $q_{1}\mapsto q_{1}-2e$.) In terms of $N_{\rm tot}$ and $\delta N$, the energy \eqref{ECij} takes the form
\begin{align}
&H_{C}=\frac{1}{2C_{\rm eff}}(2e\delta N-q_{\rm ind})^{2}+{\rm constant},\label{EdeltaQ}\\
&C_{\rm eff}=\frac{C_{11}C_{22}-C_{12}^{2}}{2C_{12}+C_{11}+C_{22}},\label{Ceffdef}
\end{align}
where the constant term is independent of $\delta N$. The induced charge $q_{\rm ind}$ is given by
\begin{align}
q_{\rm ind}={}&(\tfrac{1}{2}-f_{C})(q_{1}-en_{1})-(\tfrac{1}{2}+f_{C})(q_{2}-en_{2})\nonumber\\
&+2ef_{C}N_{\rm tot},\label{qinddef}\\
f_{C}={}&\frac{(C_{11}-C_{22})/2}{2C_{12}+C_{11}+C_{22}}.\label{fCdef}
\end{align}

We see that $q_{\rm ind}$ depends on the total number of Cooper pairs $N_{\rm tot}$, if $C_{11}\neq C_{22}$. As a consequence, when external charges are brought onto the island, the value of $q_{\rm ind}$ changes by an unknown amount, in general unequal to a multiple of $2e$. This prevents a determination of the fermion parity of the external charges.

In order to use the differential Cooper pair box as a parity meter, the Majorana fermions should be transported from one island to the other. This changes $n_{1}\mapsto n_{1}+n_{M}$, $n_{2}\mapsto n_{2}-n_{M}$, hence $q_{\rm ind}\mapsto q_{\rm ind}+en_{M}$ --- at constant $N_{\rm tot}$. That is the method described in the text.


\begin{thebibliography}{99}
\bibitem{note1} One qubit can be read out by fusion and detection of quasiparticle excitations, without any coupling to the charge and hence without any sensitivity to charge noise. Coupling to charge is necessary for the joint read out of two qubits. This follows from the fact that the fermion parity operator of two qubits (which determines its state) is the product of four Majorana operators, while the quasiparticle Hamiltonian is only quadratic in the Majorana operators.
\end{thebibliography}
\end{document}